\documentstyle[12pt]{article}

\makeatletter

\newskip\eqnshortskip
\eqnshortskip\abovedisplayshortskip
\advance\eqnshortskip-\abovedisplayskip

\if@twoside
\else \oddsidemargin 00pt \evensidemargin 00pt
\fi
\def\section{\@startsection {section}{1}{\z@}{+6.0ex plus +1ex minus
 +.2ex}{2.8ex plus .2ex}{\large\boldmath\bf}}
\def\subsection{\@startsection {subsection}{2}{\z@}{+3.0ex plus +1ex
minus +.2ex}{2.3ex plus .2ex}{\normalsize\boldmath\bf}}
\def\subsubsection{\@startsection{subsubsection}{3}{\z@}{+2.5ex plus
+1ex minus +.2ex}{1.5ex plus .2ex}{\normalsize\bf}}

\def\theequation{\thesection.\arabic{equation}}
\@addtoreset{equation}{section}

\def\appendix{\par
 \setcounter{section}{0} \setcounter{subsection}{0}
 \def\thesection{\Alph{section}}}

\let\@eqnsel = \hfil

\def\eqnarray{\stepcounter{equation}\let\@currentlabel=\theequation
\global\@eqnswtrue
\global\@eqcnt\z@\tabskip\@centering\let\\=\@eqncr
$$\halign to \displaywidth\bgroup\hskip\@centering
  $\displaystyle\tabskip\z@{##}$\@eqnsel&\global\@eqcnt\@ne
  \hskip 2\arraycolsep \hfil${##}$\hfil
  &\global\@eqcnt\tw@ \hskip 2\arraycolsep $\displaystyle\tabskip\z@{##}$\hfil
   \tabskip\@centering&\llap{##}\tabskip\z@\cr}

\expandafter\ifx\csname mathrm\endcsname\relax\def\mathrm#1{{\rm #1}}\fi
\@ifundefined{mathrm}{\def\mathrm#1{{\rm #1}}}{\relax}

\makeatother

\def\beq{\begin{equation}}
\def\eeq{\end{equation}}
\def\beqar{\begin{eqnarray}}
\def\eeqar{\end{eqnarray}}
\def\bma{\begin{displaymath}}
\def\ema{\end{displaymath}}
\def\barr#1{\begin{array}{#1}}
\def\earr{\end{array}}
\def\bit{\begin{itemize}}
\def\eit{\end{itemize}}
\def\bfi{\begin{figure}}
\def\efi{\end{figure}}
\def\btab{\begin{table}}
\def\etab{\end{table}}
\def\bce{\begin{center}}
\def\ece{\end{center}}
\def\nn{\nonumber}
\def\disp{\displaystyle}
\def\text{\textstyle}

\def\refeq#1{\mbox{(\ref{#1})}}
\def\refeqs#1{\mbox{(\ref{#1})}}
\def\reffi#1{\mbox{Fig.~\ref{#1}}}

\def\refta#1{\mbox{Tab.~\ref{#1}}}
\def\refta#1{\mbox{Table~\ref{#1}}}
\def\reftas#1{\mbox{Tables~\ref{#1}}}

\def\mathswitch#1{\relax\ifmmode#1\else$#1$\fi}
\newcommand{\mathswitchr}[1]
{\relax\ifmmode{\mathrm{#1}}\else$\mathrm{#1}$\fi}

\renewcommand{\O}{{\cal{O}}}

\newcommand{\GeV}{\unskip\,\mathrm{GeV}}

\newcommand{\PW}{\mathswitchr W}
\newcommand{\PZ}{\mathswitchr Z}

\newcommand{\PH}{\mathswitchr H}

\newcommand{\Pd}{\mathswitchr d}

\newcommand{\Ps}{\mathswitchr s}

\newcommand{\Pb}{\mathswitchr b}
\newcommand{\Pbbar}{\mathswitch {\bar{\Pb}}}
\newcommand{\Pt}{\mathswitchr t}

\newcommand{\MW}{\mathswitch {M_\PW}}

\newcommand{\MZ}{\mathswitch {M_\PZ}}
\newcommand{\MH}{\mathswitch {M_\PH}}

\newcommand{\Mb}{\mathswitch {m_\Pb}}
\newcommand{\Mt}{\mathswitch {m_\Pt}}

\newcommand{\sw}{\mathswitch {s_\PW}}
\newcommand{\cw}{\mathswitch {c_\PW}}
\newcommand{\swbar}{\mathswitch {\bar s_\PW}}
\newcommand{\cwbar}{\mathswitch {\bar c_\PW}}

\newcommand{\GF}{\mathswitch {G_\mu}}

\newcommand{\de}{\delta}
\newcommand{\ga}{\gamma}

\catcode`\@=11
\def\slash{\mathpalette\make@slash}
\def\make@slash#1#2{\setbox\z@\hbox{$#1#2$}%
  \hbox to 0pt{\hss$#1/$\hss\kern-\wd0}\box0}

\newcommand{\ps}{p\hspace{-0.42em}/}

\def\siml{\mathrel{\mathpalette\@versim<}} 
\def\simg{\mathrel{\mathpalette\@versim>}} 
\def\@versim#1#2{\lower2.5\p@\vbox{\baselineskip\z@skip\lineskip-.2\p@
    \ialign{$\m@th#1\hfil##\hfil$\crcr#2\crcr\sim\crcr}}}

\def\Zbb{\PZ\Pb\Pbbar}
\def\irr{\mathrm{irr}}
\def\ferm{\mathrm{ferm}}
\def\oneloop{\mathrm{1\dash loop}}
\def\twoloop{\mathrm{2\dash loop}}

\def\oneloop{\mathrm{1-loop}}
\def\twoloop{\mathrm{2-loop}}
\def\oneloop{\mbox{\scriptsize 1-loop}}
\def\twoloop{\mbox{\scriptsize 2-loop}}

\begin{document}
\thispagestyle{empty}
\null
\hfill CERN-TH.6874/93
\vskip 1cm
\vfil
\begin{center}
{\Large\bf\boldmath{$\O(\GF^2\Mt^4)$} Electroweak Radiative Corrections\\
to the \boldmath{$\PZ\Pb\Pbbar$} Vertex \par} \vskip 2.5em
{\large
{\sc A.\ Denner$^1$, W.~Hollik$^{2,3}$ and B.~Lampe$^3$} \\[10ex]
\parbox{10cm}{\normalsize
{\it$^1$ Theory Division, CERN, Geneva, Switzerland} \\[1ex]
{\it$^2$ Fakult\"at f\"ur Physik, Universit\"at Bielefeld, Germany}
\\[1ex]
{\it$^3$ Max-Planck-Institut f\"ur Physik, Munich, Germany}
}
\par} \vskip 1em
\end{center} \par
\vskip 4cm
\vfil
{\bf Abstract} \par
The leading heavy-top two-loop corrections to the \Zbb~vertex are
determined from a direct evaluation of the corresponding Feynman
diagrams in the large \Mt~limit.
The leading one-loop top-mass effect is enhanced by
$[1+\GF\Mt^2(9-\pi^2/3)/(8\pi^2\sqrt{2})]$.
Our calculation confirms a recent result of Barbieri et al.
\par
\vskip 2cm
\noindent CERN-TH.6874/93 \par
\vskip .15mm
\noindent April 1993 \par
\null
\setcounter{page}{0}
\clearpage

\def\gl{\alpha\Mt^2/(\pi\MW^2)}
\def\gl{\alpha\Mt^2}
\let\zbb\Zbb
\newcommand{\Gmt}{G_{\mu} \Mt^2}
\newcommand{\Gmtt}{\O(G_{\mu}^2 \Mt^4)}

\section{Introduction}

With the steadily increasing accuracy of the LEP data on the
electroweak properties of the \PZ~boson, the requirement for precise
and reliable theoretical predictions  from the Standard Model
makes the discussion of two-loop electroweak corrections
an unavoidable task. The \MZ--\MW\ correlation and the \PZ-boson
decay widths into the known light fermions contain significant
indirect effects resulting from the virtual presence of the top quark
in the loop contributions, which can be classified as follows:
\begin{itemize}
\item[(i)]
universal effects due to the large mass splitting between the \Pt~and
\Pb~quark. Their dominating source is the $\rho$-parameter, which
picks up a contribution \cite{Ve77}
$$
 \Delta\rho = 3\frac{\Gmt}{8\pi^2\sqrt{2}}
$$
at the one-loop level.
\item[(ii)]
non-universal effects in the $\zbb$ vertex,
which also involve a term $\sim \Gmt$ as the leading contribution at the
one-loop level \cite{Ak86,Be88,Be91}.
\end{itemize}
It has been known already for quite some time \cite{Co89}
that for obtaining an adequate precision on the theoretical side
the leading next order contributions $\sim (\Gmt)^2$
have to be included. This covers the proper resummation of the
one-loop contribution to the $\rho$-parameter (yielding the
1-particle reducible higher order terms)
as well as the explicit calculation of the 1-particle irreducible
(1PI) two-loop corrections to $\rho$ and to the $\zbb$ on-shell vertex
function, both of which involve terms of $\Gmtt$ in their
leading structure.

The first computation of the 1PI contribution of $\Gmtt$ to the
$\rho$-parameter has been performed in \cite{Bi87} in the
approximation of a small Higgs mass $M_H\ll m_t$.     The
generalization to arbitrary Higgs masses was done  recently
in \cite{Ba92}, where also the leading two-loop term
for the $\zbb$ vertex was given.
 The results of \cite{Ba92} were
obtained by exploiting the Ward identities of the Standard Model
Lagrangian with the gauge interactions switched off, which relate
$\rho$ and the $\zbb$ vertex corrections of $\Gmtt$ to the
field and Yukawa vertex
renormalization constants for the system of the
 unphysical Higgs fields and the $(\Pt,\Pb)$ doublet.
The result of \cite{Bi87} for the $\rho$-parameter
 is confirmed by \cite{Ba92}, taking
the limit $\MH/\Mt \rightarrow 0$.
For the $\zbb$ vertex, no further calculation has been available so
far.

In this paper we present an independent calculation of the two-loop
corrections of $\Gmtt$ to the $\zbb$ vertex.
Differently from \cite{Ba92}, our approach consists of
the direct evaluation of the $\zbb$ vertex diagrams contributing
in the considered approximation, thus allowing a verification of the
result in \cite{Ba92} by an independent method.
As a first step, we have performed the  calculation for the case
of a small Higgs mass with respect to $m_t$, where the diagrams
can be evaluated in the approximation with $m_t$ as the only mass
parameter. Our result confirms the corresponding
result of \cite{Ba92} in the limit of a light Higgs.

Section 2 contains the analytic expressions for the two-loop vertex
integrals contributing in $\Gmtt$. In section 3 we discuss the
two-loop renormalization on the basis of the on-shell scheme and provide
the two-loop contributions
following from the counter-term
structure of the $\zbb$ vertex. Thereby, the on-shell scheme is
applied in the version where wave-function renormalization
for external particles does not occur.
Hence, the resulting vertex
obtained by combining the expressions of sections 2 and 3,
 which is given and discussed in section 4,
simultaneously represents the amplitude for the decay
$\PZ\to\Pb\bar{\Pb}$ (up to the wave functions for the external
particles). It can thus be immediately transferred into the partial
width for the $\PZ\to\Pb\bar{\Pb}$ decay.

\section{Calculation of two-loop \protect\boldmath{$\O((\gl)^2)$}
contributions\protect\\ to the \protect\boldmath{$\PZ\Pb\Pbbar$} vertex}
\label{SEcalcul}

We are interested in the corrections proportional to $(\gl)^2$, i.e.\ the
leading two-loop corrections in an expansion for large \Mt.
Since for \Mt\ larger than all other scales the two-loop vertex integrals
behave like a constant or decrease with \Mt, only the diagrams with four
Yukawa couplings proportional to \Mt\ contribute.
This implies that the external
\Pb~quarks couple to a top quark and a charged Higgs ghost $\phi$, and
consequently that all contributions $\propto (\gl)^2$ are purely
left-handed just as for the corresponding one-loop corrections.
The relevant diagrams are shown in \reffi{FIverdia}.
\bfi
\begin{center}
\begin{picture}(15,11.3)
\put(-0.1,-3.9){\includegraphics{twolvert.ps}}
\end{picture}
\end{center}
\caption{Two-loop and one-loop diagrams contributing to the \Zbb~vertex}
\label{FIverdia}
\efi

For the calculation of the leading term in an expansion for large \Mt, we
can neglect the masses of all other particles ($\MH=\MW=\MZ=\Mb=0$) and
all external momenta ($p_\Pb=p_{\Pbbar}=0$) in the loop integrals.
As a consequence, we end up with only vacuum integrals.
The genuine UV and the IR singularities, which arise from putting
the masses to zero, are regularized using the na\"{\i}ve dimensional
regularization scheme with anticommuting $\ga_5$.
After renormalization all singularities cancel in the final result.

Making use of
Dirac algebra and Lorentz covariance all appearing two-loop diagrams
can easily be reduced to the following form
\def\xtil{\tilde{x}}
\beq
\ga_\mu(1-\ga_5) {\xtil}^2 \sum_i c_i G_i,
\eeq
where
\beq
\xtil=\frac{\alpha}{4\pi}\frac{1}{4\sw^2}\frac{\Mt^2}{\MW^2}.
\eeq
The coefficients $c_i$ depend  on the dimension $n$ and on the coupling
constants of the diagram under consideration.
The two-loop integrals $G_i$ are defined as
\beqar
G(a,b,c,d,e,i,j,k) &=& \int\frac{d^nq_1}{(2\pi\mu)^{n-4}i\pi^2}
\frac{d^nq_2}{(2\pi\mu)^{n-4}i\pi^2} \nn\\
&&\times\frac{(q_2^2)^i(q_1q_2)^j(q_1^2)^k}
{(q_2^2)^a(q_2^2-\Mt^2)^b(q_1-q_2)^{2c}(q_1^2-\Mt^2)^d(q_1^2)^e}.
\eeqar
The scale parameter of dimensional regularization is denoted by $\mu$.

Using standard reduction techniques the $G$'s can be reduced to the
following two types of scalar two-loop vacuum integrals:
\beqar
I_1(b,c,d) &=& \int\frac{d^nq_1}{(2\pi\mu)^{n-4}i\pi^2}
\frac{d^nq_2}{(2\pi\mu)^{n-4}i\pi^2}
\frac{1}{(q_2^2-\Mt^2)^b(q_1-q_2)^{2c}(q_1^2-\Mt^2)^d},\nn\\[1ex]
I_2(a,c,d) &=& \int\frac{d^nq_1}{(2\pi\mu)^{n-4}i\pi^2}
\frac{d^nq_2}{(2\pi\mu)^{n-4}i\pi^2}
\frac{1}{(q_2^2)^a(q_1-q_2)^{2c}(q_1^2-\Mt^2)^d},
\eeqar
which correspond to diagrams with three chains of propagators with
the same masses $m=\Mt$ or $m=0$. These integrals can be explicitly
evaluated to be (see e.g.\ \cite{Scharf})
\beqar
I_1(b,c,d) &=& (-1)^{b+c+d}\biggl(\frac{4\pi\mu^2}{\Mt^2}\biggr)^{4-n}
(\Mt^2)^{4-b-c-d} \nn\\*
&&\times\frac{\Gamma(b+c+d-n)\Gamma(b+c-n/2)\Gamma(d+c-n/2)\Gamma(n/2-c)}
{\Gamma(b+2c+d-n)\Gamma(b)\Gamma(d)\Gamma(n/2)},\nn\\[1ex]
I_2(a,c,d) &=& (-1)^{a+c+d}\biggl(\frac{4\pi\mu^2}{\Mt^2}\biggr)^{4-n}
(\Mt^2)^{4-a-c-d} \nn\\
&&\times\frac{\Gamma(a+c+d-n)\Gamma(n/2-a)\Gamma(n/2-c)\Gamma(a+c-n/2)}
{\Gamma(a)\Gamma(c)\Gamma(d)\Gamma(n/2)}.
\eeqar

With the help of the functional equation for the $\Gamma$ function, these
expressions can be reduced to the following basic integrals:
\beqar
I_1(1,0,1) &=& \biggl(\frac{4\pi\mu^2}{\Mt^2}\biggr)^{4-n}
(\Mt^2)^{2}\,[\Gamma(1-n/2)]^2, \nn\\
I_2(1,1,1) &=& \biggl(\frac{4\pi\mu^2}{\Mt^2}\biggr)^{4-n}
\Mt^2 \,\Gamma(1-n/2)\Gamma(n/2-1)\Gamma(3-n).
\eeqar
While $I_2(1,1,1)$ is a genuine two-loop integral, $I_1(1,0,1)$ is the
square of a one-loop integral.
After splitting off the common global 
factor
\def\FMS{F}
\bma
\FMS^2=\biggl(\frac{4\pi\mu^2}{\Mt^2}\biggr)^{4-n}[\Gamma(n/2-1)]^{-2},
\ema
which becomes equal to one in the limit $n\to4$, an
expansion about $n=4$ yields ($2\de = 4-n$)
\vspace{\eqnshortskip}
\beqar \label{I1I2}
I_1(1,0,1) &=& \FMS^2\,(\Mt^2)^{2}\,
\biggl(\frac{1}{\de^2}+\frac{2}{\de}+3+\frac{1}{3}\pi^2\biggr),\nn\\
I_2(1,1,1) &=& \FMS^2 \,\Mt^2\,
\biggl(\frac{1}{2\de^2}+\frac{3}{2\de}+\frac{7}{2}+\frac{1}{3}\pi^2\biggr).
\eeqar

Using the reduction algorithm and the explicit results \refeq{I1I2} we
obtain for each two-loop ($l=2$) or one-loop ($l=1$)
diagram a result of the form
\beq \label{defAB}
ie\ga_\mu\frac{1-\ga_5}{2}x^l\FMS^l
\left[A\frac{1}{2\sw\cw} + B\frac{\cw}{\sw}\right].
\eeq
The explicit expressions $A$ and $B$ for the diagrams of \reffi{FIverdia}
are listed in \refta{TAverres}.
We list all diagrams without neutral Higgs bosons separately but
give only the sums of the corresponding diagrams with internal
\PH\ and $\chi$ fields. The last four two-loop diagrams with the internal
fermion loop are denoted in the following as  `fermionic' (\ferm).

\def\disp{\displaystyle\rule[-4mm]{0mm}{10.5mm}}
\begin{table}
\begin{center}
$$\barr{l@{\qquad}c@{\qquad}c}
\hline
\mathrm{Diagram(s)} & A
& B 
\\ \hline
a), \PH+\chi &
\disp
\frac{1}{2\de^2} + \frac{3}{4\de} - \frac{81}{8} + \frac{4}{3}\pi^2  &
\disp
-\frac{1}{3\de^2} - \frac{11}{6\de} + \frac{3}{4} - \frac{4}{9}\pi^2  \\
b), \PH+\chi &
\disp
- \frac{4}{\de} + 12 - \frac{4}{3}\pi^2  &
\disp
\frac{2}{3\de^2} + \frac{5}{3\de} - \frac{9}{2} + \frac{8}{9}\pi^2  \\
c), \PH+\chi &
\disp
 \frac{1}{2\de^2} - \frac{1}{4\de} - \frac{45}{8} + \frac{2}{3}\pi^2  &
\disp
-\frac{1}{2\de^2} + \frac{1}{4\de} + \frac{45}{8} - \frac{2}{3}\pi^2  \\
d), \PH+\chi &
\disp
-\frac{1}{2\de^2} + \frac{5}{4\de} + \frac{57}{8} - \frac{2}{3}\pi^2  &
0 \\
\hline
a), \phi &
\disp
-\frac{1}{2\de^2} - \frac{3}{4\de} - \frac{7}{8} - \frac{1}{3}\pi^2  &
\disp
\frac{1}{6\de^2} + \frac{1}{4\de} + \frac{7}{24} + \frac{1}{9}\pi^2  \\
b), \phi &
\disp
-\frac{2}{\de} - 2  &
\disp
\frac{2}{3\de^2} + \frac{1}{\de} + \frac{7}{6} + \frac{4}{9}\pi^2  \\
c), \phi &
\disp
 \frac{1}{2\de^2} + \frac{3}{4\de} + \frac{7}{8} + \frac{1}{3}\pi^2  &
\disp
-\frac{1}{2\de^2} - \frac{3}{4\de} - \frac{7}{8} - \frac{1}{3}\pi^2  \\
d), \phi &
\disp
 \frac{1}{2\de^2} + \frac{3}{4\de} + \frac{7}{8} + \frac{1}{3}\pi^2  &
\disp
-\frac{1}{2\de^2} - \frac{3}{4\de} - \frac{7}{8} - \frac{1}{3}\pi^2  \\
\hline
e), \ferm &
\disp
 \frac{24}{\de^2} + \frac{60}{\de} + 108 + 8\pi^2  &
\disp
-\frac{14}{\de^2} - \frac{31}{\de} - \frac{111}{2} - \frac{16}{3}\pi^2 \\
f), \ferm&
\disp
-\frac{18}{\de^2} - \frac{39}{\de} - \frac{123}{2} - 8\pi^2  &
\disp
 \frac{18}{\de^2} + \frac{39}{\de} + \frac{123}{2} + 8\pi^2  \\
g), \ferm &
\disp
-\frac{3}{\de^2} - \frac{15}{2\de} - \frac{87}{4}  &
\disp
 \frac{1}{\de^2} + \frac{5}{2\de} + \frac{29}{4} \\
h), \ferm &
-6 &
\disp
\frac{2}{\de^2} + \frac{5}{\de} + \frac{29}{2} \\
\hline
i), \mbox{1-loop} &
 2 + 2\de & \disp
 - \frac{2}{3\de} - 1 -\frac{7}{6}\de - \frac{1}{9}\pi^2\de\\
j), \mbox{1-loop} & \disp
 - \frac{1}{\de} - \frac{3}{2} - \frac{7}{4}\de - \frac{1}{6}\pi^2\de &
\disp
   \frac{1}{\de} + \frac{3}{2} + \frac{7}{4}\de + \frac{1}{6}\pi^2\de \\
\hline
\earr$$
\end{center}
\caption{Explicit results for the two-loop vertex diagrams}
\label{TAverres}
\end{table}

\section{Renormalization and counter-term contributions}
\label{SEren}

We work in the on-shell renormalization scheme with $e=\sqrt{4\pi\alpha}$
and the particle masses as fundamental renormalized parameters and
make use of the shorthands
\beq
\cw=\frac{\MW}{\MZ}, \qquad \sw=\sqrt{1-\cw^2}.
\eeq
In a second step, we relate these parameters to the Fermi constant \GF.

The renormalized amputated vertex function including all contributions up
to two loops can be written as
\beqar
\Gamma^{\Zbb}_{\mu,2}&=& Z_{\Pb,L}\sqrt{Z_\PZ}\biggl[
\Gamma^{\Zbb}_{\mu,0}(e+\de e,\cw^2+\de\cw^2,\sw^2+\de\sw^2)\nn\\
&&{}+\Delta\Gamma^{\Zbb}_{\mu,\oneloop}(e+\de e,\cw^2+\de\cw^2,
\sw^2+\de\sw^2,\Mt+\de\Mt,M_\phi^2+\de M_\phi^2,\MW^2+\de\MW^2)
\biggr]\nn\\
&&{}+\Delta\Gamma^{\Zbb}_{\mu,\twoloop},
\eeqar
where $\Gamma^{\Zbb}_{\mu,0}$ is the bare vertex function,
$\Delta\Gamma^{\Zbb}_{\mu,\oneloop}$ and
$\Delta\Gamma^{\Zbb}_{\mu,\twoloop}$
the unrenormalized one-loop and two-loop contributions. The unrenormalized
quantities depend on the bare parameters, which have been split into
renormalized ones and counter terms. Finally $Z_{\Pb,L}$ and $Z_\PZ$
are the field-renormalization constants for the external left-handed
\Pb-quark and \PZ-boson fields.

In the order $(\gl)^2$ there are no contributions to $\de e$ and
$\de Z_\PZ = Z_\PZ -1$. Thus the only genuine two-loop counter terms
are due to $Z_{\Pb,L} = 1 + \de Z^{(1)}_{\Pb,L} + \de Z^{(2)}_{\Pb,L}$
and $\de\cw^2$, $\de\sw^2$.
The latter are finite and represent the one- and two-loop effects
of the $\rho$-parameter \cite{Co89,Bi87,Ba92}
\beq
\frac{\de\cw^2}{\cw^2} = \frac{1}{\rho}-1, \qquad
\frac{\de\sw^2}{\sw^2} = -\frac{\cw^2}{\sw^2}\frac{\de\cw^2}{\cw^2},
\eeq
where
\beq \label{rho}
\rho = \frac{1}{1-\de\rho_{\irr}},
\eeq
with
\beq \label{drho}
\de\rho_{\irr} = 3[x - (2\pi^2-19)x^2]
\eeq
for $\MH\ll\Mt$ and
\beq \label{x}
x= \frac{\GF\Mt^2}{8\pi^2\sqrt{2}}.
\eeq
These are the universal effects, which are also present for \Pd\
or \Ps~quarks.
They can simply be taken into account by replacing $\sw$ and $\cw$
everywhere by
\beq
\cwbar = \sqrt{\cw^2 + \de\cw^2}, \qquad \swbar = \sqrt{\sw^2 + \de\sw^2}
\eeq
\def\xbar{\bar x}
and will not be discussed any further.

Thus we find to order $(\gl)^2$:
\beqar
\Gamma^{\Zbb}_{\mu,2}&=& (1 + \de Z^{(1)}_{\Pb,L} + \de Z^{(2)}_{\Pb,L})
\Gamma^{\Zbb}_{\mu,0}(e,\cwbar^2,\swbar^2)\nn\\
&&{}+(1 + \de Z^{(1)}_{\Pb,L})
\Delta\Gamma^{\Zbb}_{\mu,\oneloop}(e,\cwbar^2,\swbar^2,
\Mt+\de\Mt,M_\phi^2+\de M_\phi^2,\MW^2+\de\MW^2)\nn\\
&&{}+\Delta\Gamma^{\Zbb}_{\mu,\twoloop}.
\eeqar

The only non-universal two-loop counter term is $\de Z^{(2)}_{\Pb,L}$.
{}From the on-shell renormalization condition, which requires residue
unity for the renormalized \Pb~propagator, one obtains
\beq
Z_{\Pb,L} = \frac{1}{1+\Sigma_L^{\Pb\Pbbar}(0)}
\eeq
and thus
\beq \label{Z2b}
\de Z^{(2)}_\Pb = -\Sigma_{L,\twoloop}^{\Pb\Pbbar}(0)
+[\Sigma_{L,\oneloop}^{\Pb\Pbbar}(0)]^2.
\eeq
Here $\Sigma_L^{\Pb\Pbbar}$ is the left-handed scalar coefficient of the
\Pb-quark self-energy
\beq
\Sigma^{\Pb\Pbbar}(p)=\ps\frac{1-\ga_5}{2}\Sigma_{L}^{\Pb\Pbbar}(p^2).
\eeq
The corresponding right-handed part vanishes in the considered
approximation. Using the relation
\beq
\ga_\mu\frac{1-\ga_5}{2}\Sigma_L^{\Pb\Pbbar}(0)=
\left.\frac{\partial\Sigma^{\Pb\Pbbar}(p)}{\partial p^\mu}\right|_{p=0}
\eeq
we obtain $\Sigma_L^{\Pb\Pbbar}(0)$ by differentiating the self-energy
diagrams with respect to the external momentum $p$ and setting $p=0$
afterwards. Thus also $\Sigma_L^{\Pb\Pbbar}(0)$ can be calculated
from only vacuum integrals.

The relevant diagrams for $\Sigma^{\Pb\Pbbar}$ are given in
\reffi{FIZbLdia}.
\bfi
\begin{center}
\begin{picture}(16,2.5)
\put(-0.0,-8.5){\includegraphics{twolself.ps}}
\end{picture}
\end{center}
\caption{Two-loop and one-loop diagrams
contributing to $\Sigma^{\Pb\Pbbar}$}
\label{FIZbLdia}
\efi
At order $\alpha^2$, in addition to the two-loop diagrams
we must include the counter terms
originating from the one-loop renormalization of $\Sigma^{\Pb\Pbbar}$:
\beq
\hat\Sigma^{\Pb\Pbbar} = Z_\Pb\Sigma^{\Pb\Pbbar}(e,\cwbar,\swbar,
\Mt+\de\Mt,M_\phi^2+\de M_\phi^2,\MW^2+\de\MW^2).
\eeq
These are easily generated from the one-loop diagram~(c)
in \reffi{FIZbLdia}.
One way is to take the analytical result for this diagram
before evaluation of the integrals and for finite
$M_\phi^2$, replace the bare parameters by the renormalized ones
and the counter terms, multiply with $1 + \de Z_\Pb^{(1)}$, and expand up
to the first order in the counter terms.
Alternatively one generates all counter-term diagrams corresponding to
this one-loop diagram by inserting one-loop
counter terms into all possible propagators and vertices, and evaluates
these counter-term diagrams individually.

The single contributions of
$\de Z_\Pb^{(2)}$ to $A$ and $B$ (see \refeq{defAB}) are
listed in \refta{TAZbLres}.
\begin{table}
\begin{center}
 $$\barr{l@{\qquad}c@{\qquad}c}
\hline
\barr{l} \mathrm{Diagram(s)\ or}\\
\mathrm{counter~term(s)} \earr & A & B \\
\hline
a), \PH+\chi &
\disp
-\frac{1}{2\de^2} + \frac{1}{4\de} + \frac{45}{8} - \frac{2}{3}\pi^2  &
\disp
 \frac{1}{6\de^2} - \frac{1}{12\de} - \frac{15}{8} + \frac{2}{9}\pi^2  \\
\hline
a), \phi &
\disp
-\frac{1}{2\de^2} - \frac{3}{4\de} - \frac{7}{8} - \frac{1}{3}\pi^2  &
\disp
\frac{1}{6\de^2} + \frac{1}{4\de} + \frac{7}{24} + \frac{1}{9}\pi^2  \\
\hline
b), \ferm &
\disp
 \frac{21}{\de^2} + \frac{93}{2\de} + \frac{333}{4} + 8 \pi^2  &
\disp
-\frac{7}{\de^2} - \frac{31}{2\de} - \frac{111}{4} - \frac{8}{3}\pi^2 \\
\hline
\de\Mt^{\PH+\chi} &
\disp
 \frac{2}{\de^2} + \frac{7}{\de} + \frac{35}{2} + \frac{2}{3} \pi^2  &
\disp
-\frac{2}{3\de^2} - \frac{7}{3\de} - \frac{35}{6} - \frac{2}{9} \pi^2  \\
\hline
\de\Mt^{\phi} &
\disp
 \frac{1}{\de^2} + \frac{5}{2\de} + \frac{21}{4} - \frac{1}{3} \pi^2  &
\disp
-\frac{1}{3\de^2} - \frac{5}{6\de} - \frac{7}{4} + \frac{1}{9} \pi^2  \\
\hline
\de\MW^2,\de M_\phi^2 &
\disp
-\frac{18}{\de^2} - \frac{48}{\de} - \frac{171}{2} - 6 \pi^2  &
\disp
 \frac{6}{\de^2} + \frac{16}{\de} + \frac{57}{2} + 2 \pi^2  \\
\hline
\de Z_{\Pb,L}^{(1)} &
\disp
-\frac{1}{\de^2} - \frac{3}{\de} - \frac{23}{4} - \frac{1}{3} \pi^2  &
\disp
 \frac{1}{3\de^2} + \frac{1}{\de} + \frac{23}{12} + \frac{1}{9} \pi^2  \\
\hline
c), \mbox{1-loop} &
\disp
 \frac{1}{\de} + \frac{3}{2} + \frac{7}{4}\de + \frac{1}{6}\pi^2\de  &
\disp
-\frac{1}{3\de} - \frac{1}{2} - \frac{7}{12}\de - \frac{1}{18}\pi^2\de \\
\hline
\earr$$
\end{center}
\caption{Explicit results for the various contributions of
$\de Z_{\Pb,L}$}
\label{TAZbLres}
\end{table}
While the loop diagrams are characterized by their internal fields, the
counter-term contributions are separated according to the various
renormalization constants. They can be associated with the loop diagrams
as follows (see also \reffi{FIctdia}).  The counter terms
$\de\MW^2$ and $\de M_\phi^2$ originate from
`fermionic' diagrams, $\de\Mt$ gets contributions from
$\PH+\chi$ and $\phi$, $\de Z_{\Pb,L}^{(1)}$ only from $\phi$.
The contribution of $\de Z_{\Pb,L}^{(1)}$
to $\de Z_{\Pb,L}^{(2)}$ gives rise to the term
$[\Sigma_{L,\oneloop}^{\Pb\Pbbar}(0)]^2$ in \refeq{Z2b}.
As discussed above the counter terms $\de\sw^2$ and $\de\cw^2$
are absorbed in $\swbar$ and $\cwbar$.
This leads in particular to the replacement
\beq
\xtil \to \xbar =
\frac{\alpha}{4\pi}\frac{1}{4\swbar^2}\frac{\Mt^2}{\MW^2}
\eeq
in the one-loop contribution.

The one-loop counter terms are determined from the diagrams in
\reffi{FIctdia} and diagram~(c) in \reffi{FIZbLdia}.
In the on-shell scheme the $\phi$~mass counter term
is directly related to the tadpole counter term $\delta t$ as
\beq
\de M_\phi^2 = - \frac{e}{2\MW\sw} \de t
\eeq
and thus determined from the tadpole diagram. In the considered
approximation the value for $\de M_\phi^2$ turns out to be equal to
the one obtained from an on-shell condition to the $\phi$ self-energy.
Explicitly we obtain
\beqar \label{cou}
\de\MW^2 &=& -6x\MW^2
\biggl(\frac{1}{\de} + \frac{1}{2} + \frac{1}{4}\de + \frac{\pi^2}{6}\de
\biggr), \nn \\
\de M_\phi^2 &=& -12x\Mt^2
\biggl(\frac{1}{\de} + 1 + \de + \frac{\pi^2}{6}\de \biggr), \nn \\
\de\Mt^{\PH+\chi} &=&  x\Mt
\biggl(\frac{1}{\de} + 3 + 7\de + \frac{\pi^2}{6}\de \biggr), \nn \\
\de\Mt^{\phi} &=& x\Mt
\biggl(\frac{1}{2\de} + 1 + 2\de - \frac{\pi^2}{4}\de\biggr),
\nn \\
\de Z_{\Pb,L}^{(1)} &=& -x
\biggl(\frac{1}{\de} + \frac{3}{2} + \frac{7}{4}\de
+ \frac{\pi^2}{6}\de\biggr),
\eeqar
where we have split $\de\Mt$ into the contributions from the neutral
Higgses \PH\ and $\chi$, and from the charged Higgs $\phi$.
It is essential to take into account the terms proportional to $\de$
in \refeq{cou}.
Note that $\de\Mt$ is the only quantity that involves loop
integrals with non-vanishing external momenta.

The one-loop counter terms to the one-loop \Zbb-vertex diagrams~(i)
and (j) in \reffi{FIverdia} are obtained in exactly the same way as
the counter terms to the \Pb\Pbbar~self-energy.
\bfi
\begin{center}
\begin{picture}(10.7,2)
\put(- 0.0,-8.7){\includegraphics{twolcoun.ps}}
\end{picture}
\end{center}
\caption{Diagrams contributing to the one-loop counter terms}
\label{FIctdia}
\efi
The corresponding counter-term contributions to $A$ and $B$ are listed in
\refta{TAvctres}. They all result from products of one-loop diagrams.
\begin{table}
\begin{center}
 $$\barr{l@{\qquad}c@{\qquad}c}
\hline
\mathrm{Counter~term(s)}  & A & B \\
\hline
\de\Mt^{\PH+\chi} &
\disp
-\frac{2}{\de^2} - \frac{3}{\de} - \frac{11}{2} - \frac{2}{3} \pi^2  &
\disp
 \frac{2}{3\de^2} + \frac{7}{3\de} + \frac{35}{6} + \frac{2}{9} \pi^2  \\
\hline
\de\Mt^{\phi} &
\disp
-\frac{1}{\de^2} - \frac{1}{2\de} - \frac{5}{4} + \frac{1}{3} \pi^2  &
\disp
 \frac{1}{3\de^2} + \frac{5}{6\de} + \frac{7}{4} - \frac{1}{9} \pi^2  \\
\hline
\de\MW^2,\de M_\phi^2 &
\disp
-\frac{6}{\de^2} - \frac{12}{\de} - \frac{33}{2} - 2 \pi^2  &
\disp
-\frac{6}{\de^2} - \frac{16}{\de} - \frac{57}{2} - 2 \pi^2  \\
\hline
\de Z_{\Pb,L}^{(1)} & \disp
 \frac{1}{\de^2} + \frac{1}{\de} + \frac{3}{4} + \frac{1}{3} \pi^2  &
\disp
-\frac{1}{3\de^2} - \frac{1}{\de} - \frac{23}{12} - \frac{1}{9} \pi^2  \\
\hline
\earr$$
\end{center}
\caption{Explicit results for the counter-term contributions to the
\Zbb~vertex}
\label{TAvctres}
\end{table}

The vertex
diagrams involving $\de M_\phi^2$ contain artificial IR singularities,
which
cancel against similar contributions in the corresponding two-loop vertex
diagrams~(e) and (f).

\section{Results}
\label{SEres}

The \Zbb~vertex including corrections $\propto (\gl)^n$ has the general
form
\beq
\Gamma^{\Zbb}_{\mu,2} = -i\frac{e}{2\swbar\cwbar} \ga_\mu
   \left[\frac{1-\ga_5}{2} (1+\tau) - \frac{2}{3} \swbar^2\right],
\eeq
where $\swbar$ and $\cwbar$ include the universal corrections associated
with the $\rho$-parameter and the quantity
\beq \label{tau}
\tau = \tau^{(1)} + \tau^{(2)} + \ldots
\eeq
summarizes the leading non-universal corrections to the \Zbb~vertex.
At one loop it is given by
\beq \label{tau1}
\tau^{(1)} = -2\xbar,
\eeq
at two loops we find
\beq \label{tau2}
\tau^{(2)} = -2\xbar^2\biggl(9-\frac{\pi^2}{3}\biggr) ,
\eeq
in accordance with the result of Barbieri et al.\ \cite{Ba92}.

The quantity $\tau^{(2)}$ gets the following individual contributions
\beqar
\tau^{(2)}_{\PH+\chi} &=&
-2\xbar^2\biggl(\frac{1}{\de}+\frac{21}{2}-\frac{\pi^2}{3}\biggr), \nn\\
\tau^{(2)}_{\phi} &=&
-2\xbar^2, \nn\\
\tau^{(2)}_{Z_{\Pb,L}^{(1)}} &=&
-2\xbar^2\biggl(-\frac{1}{\de}-\frac{5}{2}\biggr), \nn\\
\tau^{(2)}_{\ferm} &=& 0,
\eeqar
where $\tau^{(2)}_{\PH+\chi}$ includes the contributions from
$\de\Mt^{\PH+\chi}$,
$\tau^{(2)}_{\phi}$ those from $\de\Mt^{\phi}$, and
$\tau^{(2)}_{\ferm}$ those from $\de\MW^2$ and $\de M_{\phi}^2$.
Particularly interesting is the fact that the contributions of the
fermion loops cancel exactly, apart from the universal contributions
entering $\swbar^2$, $\cwbar^2$.
This is also valid for additional heavy fermion doublets
with large mass splitting.
As a consequence of the conservation of the vector current, all terms
proportional to $Q_b\cw\!/\!\sw$ drop out and the final result is
proportional to $I^3_\Pb/(2\sw\cw)$.
This holds separately for the sums of all contributions denoted by
$\PH+\chi$, $\phi$ and $\ferm$ in \reftas{TAverres} and \ref{TAZbLres}
and $\de\Mt^{\PH+\chi}$, $\de\Mt^\phi$, ($\de\MW^2,\de M_\phi^2$) and
$\de Z_{\Pb,L}^{(1)}$ in \reftas{TAZbLres} and \ref{TAvctres}.

The effects of the running of $\alpha$ can easily be included by
replacing $\alpha$ by $\bar\alpha=\alpha(\MZ)$ everywhere.
Using subsequently
\beq \label{GF}
\GF \approx \frac{\pi\bar\alpha}{\sqrt{2}\swbar^2\MW^2}
\approx \frac{\pi\bar\alpha}{\sqrt{2}\swbar^2\cwbar^2\MZ^2\rho},
\eeq
which holds apart from non-leading corrections,
we may replace $\bar\alpha/\swbar^2$ by $\GF$.
This turns $\xbar$ into $x$, which was already used in \refeqs{drho} and
\refeq{x} for $\de\rho_{\irr}$, and we obtain for the partial width,
including all corrections $\propto (\Gmt)^n$
\beqar \label{GZ}
\Gamma^{\PZ\to\Pb\Pbbar} &=& \MZ\frac{\GF\MZ^2\rho}{8\sqrt{2}\pi}
\biggl[(1+\tau)^2+\biggl(1-\frac{4}{3}\swbar^2+\tau\biggr)^2 \biggr]
 &=& \MZ\frac{\GF\MZ^2\rho_\Pb}{8\sqrt{2}\pi}
\biggl[1 + \biggl(1-\frac{4}{3}\swbar^2\kappa_\Pb\biggr)^2 \biggr]
\eeqar
with
\beq
\rho_\Pb = \rho(1+\tau)^2, \qquad \kappa_\Pb = \frac{1}{1+\tau}.
\eeq
This generalizes the result of Akhundov et al.\ \cite{Ak86} to higher
orders.
While the last two equations are valid to all orders in $\Gmt$,
so far only the first and second order irreducible contributions
to $\rho$ and $\tau$ are known.
The complete one-loop corrections can be included by appropriate
redefinitions of $\kappa_\Pb$ and $\rho_\Pb$.

Numerically we find for $\Mt=200\GeV$:
\beq
\tau^{(2)} \approx -2.0 \times 10^{-4}
\eeq
and the following contributions to
$\Gamma^{\PZ\to\Pb\Pbbar}$
\begin{tabbing}
corrections \= one-loop order $\Gmt$ from $\tau^{(1)}$ and $\rho^{(1)}$
aaaa\=  $-1.9\%$,\kill
\> one-loop order $\Gmt$ from $\rho^{(1)}$  \>  $ \phantom{-}1.80\%$,\\
\> one-loop order $\Gmt$ from $\tau^{(1)}$  \>  $-1.93\%$, \\
\> remaining non-leading one-loop      \>  $-0.65\%$, \\
\> two-loop order $(\Gmt)^2$ from $\tau^{(2)}$ and $\rho^{(2)}$
\> $-0.05\%$.
\end{tabbing}
The contributions of $\tau$ and $\rho$ were
obtained from \refeq{GZ} with $\swbar^2$ determined
from \refeq{GF} and normalized to the corresponding result for
$\tau=0$ and $\rho=1$.
The one-loop results show that the leading term gives only the order of
magnitude of the complete corrections. As the leading two-loop result is
small, the complete two-loop result is presumably also small.
Consequently the perturbative series seems to converge well and the
limits on \Mt\ derived from the one-loop results should be reliable.

Recently also the strong-electroweak
two-loop corrections of $\O(\alpha_s\Gmt)$
to the $\Zbb$ vertex have been evaluated \cite{Fl92}. Thus the
leading electroweak and QCD corrections to this vertex are now known
to the same level of accuracy as the ones to the $\rho$-parameter.

\end{document}